\documentclass[twocolumn,showpacs,preprintnumbers,aps,
prd,superscriptaddress,nofootinbib,tightenlines,floats,floatfix]{revtex4}

\usepackage{graphicx}
\usepackage{latexsym}
\usepackage{amsmath,amssymb}        
\usepackage[draft=false]{hyperref}

\begin{document}



\title{Comment on ``Tainted evidence: cosmological model selection
  versus fitting'',\\ by Eric V.~Linder and Ramon Miquel
  (astro-ph/0702542v2)} 
\author{Andrew R.~Liddle}
\affiliation{Astronomy Centre, University of Sussex, Brighton BN1 9QH,
United Kingdom}
\affiliation{Institute for Astronomy, University of Hawai`i, 2680
  Woodlawn Drive, Honolulu, HI 96822, USA}
\author{Pier Stefano Corasaniti}
\affiliation{LUTH, CNRS UMR 8102, Observatoire de Paris-Meudon, 5
  Place Jules Janssen, 
92195 Meudon Cedex, France}
\author{Martin Kunz}
\affiliation{D\'epartement de Physique Th\'eorique, Universit\'e de
  Gen\`eve, 24 quai Ernest Ansermet, CH--1211 Gen\`eve 4, Switzerland}
\author{Pia Mukherjee}
\affiliation{Astronomy Centre, University of Sussex, Brighton BN1 9QH,
United Kingdom}
\author{David Parkinson}
\affiliation{Astronomy Centre, University of Sussex, Brighton BN1 9QH,
United Kingdom}
\author{Roberto Trotta}
\affiliation{Oxford University, Astrophysics, Denys Wilkinson
  Building, Keble Road, OX1 3RH, United Kingdom}
\date{\today}
\pacs{98.80.-k \hfill astro-ph/0703285}
\preprint{astro-ph/0703285}


\begin{abstract}
In astro-ph/0702542, Linder and Miquel seek to criticize the use of
Bayesian model selection for data analysis and for survey forecasting
and design. Their discussion is based on three serious
misunderstandings of the conceptual underpinnings and application of
model-level Bayesian inference, which invalidate all their main
conclusions. Their paper includes numerous further inaccuracies,
including an erroneous calculation of the Bayesian Information
Criterion. Here we seek to set the record straight.
\end{abstract}

\maketitle


\section{Introduction}

In a recent paper, Linder and Miquel \cite{LM} have mounted a vigorous
attack on the use of model selection techniques in cosmology,
particularly with regard to interpreting (forecasting) the outcome of
(upcoming) surveys and in survey design applications. They instead
advocate a frequentist parameter-fitting technique.

Our aim in this short note is to highlight important misunderstandings
that invalidate all the main conclusions of their paper. In the
process, we give a brief self-contained discussion of the model
selection framework; for more details see
e.g.~Refs.~\cite{revs,LasHob,Trotta}. In the Appendix we highlight
some specific inaccuracies in Ref.~\cite{LM}, many of which are
consequences of the general misunderstandings outlined in the main
body of this Comment.

\section{What is Bayesian model selection?}

In Bayesian inference, model parameters are taken as random variables,
because this allows propagation of the experimental measurement errors
into self-consistent probabilistic statements about parameter
uncertainties.

The first step of Bayesian parameter estimation is the choice of a
model ($M_i$), which specifies a set of parameters ($\vec{\theta}_i$)
to be varied in fitting to the data, along with a set of prior
probability ranges $P(\vec{\theta}_i|M_i)$ for those parameters. Given
a particular set of data $D$, the likelihood $P(D|\vec{\theta}_i,M_i)$
is used to update the prior probabilities to the posterior
\begin{equation} \label{bayes}
P(\vec{\theta}_i|D,M_i) = \frac{P(D|\vec{\theta}_i,M_i) \,
P(\vec{\theta}_i|M_i)}{P(D|M_i)} \,.
\end{equation}
The posterior $P(\vec{\theta}_i|D,M_i)$ contains all the information
about the state of knowledge on the parameters $\vec{\theta}_i$ after
the arrival of the data. From this one can construct `credible
intervals', i.e. ranges encompassing say $68\%$ or $95\%$ of posterior
probability for the parameters. 

We remark that Bayesian credible intervals have a profoundly different
meaning from frequentist confidence regions, where model parameters
are not random variables but fixed unknown quantities.  The fact that
the two intervals are formally equal in the case of a Gaussian
likelihood (and flat priors, in the Bayesian scheme) is traceable to
the symmetry between the measured mean and the `true' mean entering
the Gaussian distribution. This formal equivalence can engender
considerable confusion as to the different interpretations of the
final result (for a detailed discussion see Ref.~\cite{C95}).

Bayesian model selection (or comparison) is the extension of the
parameter estimation framework to include multiple models, with
different parameter vectors and priors. Bayes theorem can be applied
again to update a prior model probability by the \emph{evidence}, also
known as the marginal likelihood of the model, which is the
normalization constant in Eq.~\eqref{bayes}
\begin{equation}
P(D|M_i) = \int P(D|\vec{\theta}_i,M_i) P(\vec{\theta}_i|M_i)
d\vec{\theta}_i \,.
\end{equation}
The evidence is the probability of the data given the model. Bayes
theorem is then used to obtain the probability of the model given
the data,
\begin{equation}
P(M_i|D) \propto P(D|M_i) \, P(M_i) \,,
\end{equation}
where $P(M_i)$ is the prior model probability. It is clear from the
above equations that the evidence is the basis of the model comparison
and is built upon the parameter estimation step.

All of the above is uncontroversial mathematics, providing a
consistent and systematic inference system for evolving probabilities
in light of experimental data. Any controversy about Bayesian methods
centres around the explicit need to state the full set of prior
information in order to do any calculation. The framework provides no
guidance as to how to do this; instead physical insight is needed to
select suitable models for comparison with data, and to assess their
initial probability and the priors on the model parameters in advance
of that comparison.  Within a model, prior parameter ranges can be
thought of as plausible regions of parameter space that are accessible
to the model. 

Bayes theorem can be thought of as a decomposition of the final result
into prior knowledge and the likelihoods measuring the information
coming from the data. Indeed, the Bayesian formalism forces us to
state explicitly which part of the result is due to our assumptions,
and which part is driven by the data. The hope or expectation, both at
the parameter level and model level, is that data will be obtained of
sufficient quality to overturn incorrect prior hypotheses.  If there
is a broad range of possible models, corresponding to different prior
choices, it will require more data to converge to a robust
conclusion. But in the Bayesian approach data will eventually
overcome prior choices; the wider the range of plausible priors, the
more data we can expect to need before a firm conclusion can be drawn.

The Bayesian evidence sets up a tension between the ability of a model
to fit the data and the prior predictiveness of the model, in a
quantitative implementation of Occam's razor. Note that that we prefer
to use the term `predictiveness' rather than `simplicity/complexity';
the former is what is actually rewarded by the evidence, and is not
necessarily directly related to, for instance, the number of
parameters. The models that do best are the ones that make specific
predictions that later turn out to fit the data well. Less predictive
models, even if they can fit the data as well, score more poorly. The
Bayesian evidence has been widely applied to cosmological problems in
recent years \cite{ev,BCK,Trotta,MPCLK,KTP,PLMP,LMPW,LasHob,PPOD}.

Some statistics have been extensively used as proxies to the actual
evidence, such as the Bayesian Information Criterion (BIC) \cite{S78,
L04}. But unlike the evidence these approximations are often biased,
and by construction disfavour models with more parameters, even when
those parameters are not constrained by the data (see IIIC for more on
this in relation to the evidence, and Ref.~\cite{L04} for a discussion
of the limitations of some information criterion based
approaches). Wherever possible the full evidence should be used.

\section{Misconceptions about model selection}

The paper of Linder and Miquel \cite{LM} launches a primarily
rhetorical attack on the model selection framework. We will argue here
that the paper contains numerous factually-incorrect statements. These
appear largely to be traceable to three fundamental misunderstandings
concerning the Bayesian framework and its applications, which we now
describe.

\subsection{Model selection does not replace parameter estimation. It
  extends it.}

Linder and Miquel appear to believe that model selection and parameter
estimation are competing techniques. This is incorrect. As described
above, model selection extends the Bayesian framework to the model
level. Within each model, parameter estimation is carried out in the
usual manner. This would include, as usual, goodness-of-fit and data
subset consistency checks.

Specifically, we see that parameter estimation corresponds to model
selection where the prior model probabilities of all but one model
have been set to zero. This seems a regressive step; one can hardly
claim that our understanding of, for instance, dark energy is so good
that we should focus on only one possible description.

From this perspective, the need to choose model priors is clearly
an advantage, not a drawback. Parameter estimation corresponds to
one particular choice of those priors. By acknowledging that other
choices are possible, a much more wide--ranging and robust
investigation of the possible outcomes of future experiments can
be made, as was done in Ref.~\cite{LMPW}.

A further advantage of model selection is that it allows one to ask
new types of question. As it subsumes parameter estimation, one can
obviously still ask about parameter confidence ranges, for instance,
either model-by-model or via Bayesian model averaging as in
Ref.~\cite{LMPW}. But one can also ask whether entire models are
excluded by data at a given strength of evidence, based on their
posterior model probability, or whether data provide support for
additional model parameters. Indeed, the current leading questions in
dark energy studies are of model selection type, viz.~is the equation
of state $w$ equal to $-1$ or is it variable?  In the latter case, is
$w$ constant or time-varying? One can also compare models that are not
nested, for instance is quintessence a better description of the data
than a modified gravity model?  Such questions are not accessible to
parameter fitting analyses and often cannot even be phrased in
frequentist terms.

An important application of model selection is survey forecasting
and design, where one assesses or constructs a survey in order to
optimize the ability to answer a particular question or questions
\cite{Trotta,MPCLK, PPOD}. The details will inevitably depend to
some extent on prior assignments, and it is of course important to
vary these within reasonable ranges. Model selection forecasting
allows optimization for a broader range of possible questions.

Linder and Miquel also claim that survey design based on model
selection is betting on the absence of structure in the possible
parameter space, apparently confusing the space of possible `true'
models with the likelihood in that space given a particular `true'
model. The opposite is true. By including several models, one can
focus attention on particular regions of parameter space that are
especially well motivated, for instance $\Lambda$CDM, or the
locations predicted by one-parameter quintessence models. In fact
it is parameter estimation that assumes that the parameter space
is a blank canvas in which each point is of equal value.

\subsection{Physical intuition and priors are the same thing!}

Linder and Miquel criticize the Bayesian methodology for giving
results that are often dependent on prior assumptions, and
simultaneously claim that it seeks to avoid, or even prevent, use
of physical intuition. Apparently they have not realised that
physical intuition and priors are the same thing! After all, where
do the models come from that we decide to compare to the data?
What decides their prior model probabilities, and the reasonable
ranges for their parameters?  This is where the physics comes in.
The mere fact that it can be difficult to put our physical
intuition in quantitative terms by selecting prior ranges and
prior probabilities is no good reason to give up the exercice.

The Bayesian model selection framework, by allowing us to specify
multiple models with both model and parameter priors, \emph{maximizes}
our chance to incorporate physical intuition into data
analysis. Linder and Miquel's claims to the contrary hold no substance
at all. From this perspective, the prior dependence in Bayesian
analysis should be viewed in a positive light, not a negative one, as
it allows different intuitions to be tested.  Bayes' theorem provides
a convenient decomposition into the parts of the conclusions that are
data-driven (the parameter and model likelihoods) and those that are
prior-driven (the physical intuition), and so one can always keep
track of the balance between those two. As long as the data cannot
decide the issue, our physical intuition influences the outcome of our
conclusions, but in the Bayesian framework we are made explicitly
aware of this situation through the need to specify an explicit
prior. There is no inference without assumptions.  As the amount and
quality of data increases, the priors become less important and the
conclusions based on our expectations are replaced by conclusions
based on actual data. This is how physics should work.

\subsection{Model selection does not act against models whose
  parameters cannot yet be measured.}

Linder and Miquel give a historical overview, titled `reality check',
which seeks to show by example that model selection techniques, if
applied in the past, would have led researchers astray. In our view
all of this section is incorrect, as we explain in detail in the
Appendix. Here we will address the reason why Linder and Miquel have
gone astray.

Their main mistake is a failure to recognize the difference between
two distinct circumstances. The first is a situation where a
phenomenon could have been discovered, but wasn't; this corresponds to
a likelihood function well localized within the prior of the relevant
parameter, but consistent with a zero value.  Model selection
statistics act against models with the extra parameter in that case
(an example being spatial curvature). The second is the situation
where observations were of insufficient power to constrain the
parameter, corresponding to a flat or nearly flat likelihood across
the prior. In this case, the contribution of the parameter factorizes
out of the evidence integral, leaving it unchanged. Therefore
\emph{Bayesian model selection does not act against parameters that
are unconstrained by existing data} (see Ref.~\cite{KTP} for a
detailed discussion).  Comparisons of such models are inconclusive,
awaiting new data.  All the examples they give purporting to show
model selection going astray are actually in the second category, and
not in the first as they say.

Their misunderstanding can be partially traced to their use of the BIC
\cite{S78}. This model selection criterion assumes that all parameters
are well measured. If this is not the case, then the BIC will exclude
models that are perfectly acceptable when using full Bayesian model
comparison, as e.g.~demonstrated in Ref.~\cite{BCK}. There it was
shown that the BIC rules out the ``kink'' parametrisation of the dark
energy equation of state, in disagreement with the full Bayesian
evidence. Indeed, in the derivation of the BIC only the scaling with
the number of data points $N$ was kept, while even in the idealized
case of linear models with Gaussian errors the overall scaling of the
log-evidence for a model with $k$ parameters is rather $k\ln(N/k)$
\cite{KTP}. Additionally there is a term that depends on the size of
the error bars relative to the size of the prior, which often
dominates. For these reasons, the BIC tends to give an unrealistically
high penalty to extra parameters, compared to the full Bayesian
evidence, if its underlying assumptions are not met. Only the evidence
is a full implementation of Bayesian model comparison.

\section{Conclusions}

We regard model selection techniques as a powerful tool for
cosmologists, both for data analysis and for survey forecasting and
design. They broaden the range of questions one can ask of present and
future observations, and can be applied in a consistent and rigorous
framework. While there remains room for debate about the relative
merits of frequentist and Bayesian approaches in cosmology, we believe
that the many demonstrable flaws of the Linder--Miquel paper prevent
it from contributing constructively to that debate.


\begin{acknowledgments}

P.S.C.\ was supported by CNRS with contract No.\ 06/311/SG, M.K.\ by
the Swiss NSF, P.M. and D.P.\ by PPARC (UK), and R.T.\ by the Royal
Astronomical Society's Sir Norman Lockyer Fellowship and by St.\
Anne's College, Oxford. We thank Eric Linder and Ramon Miquel for
discussing the content of their paper with us. A.R.L.\ additionally
thanks Gang Chen, Mark Neyrinck, Adrian Pope, and Istvan Szapudi for
discussions.
\end{acknowledgments}

\appendix

\section{Detailed critique}
In this Appendix we provided a detailed critique of some of points made
by Linder and Miquel.

\subsection{Linder and Miquel: Section III/V}

In our view, the discussion in their section III, seeking instances
from history where model selection would have misinformed, is entirely
wrong or irrelevant. Since this is the sole motivation for their
Section V, it too has no validity. We take their account paragraph by
paragraph.

a) This paragraph claims that, pre-1998, model selection would have
dismissed the now-favoured $\Lambda$CDM model. Absolutely not! Data
before that epoch were unable to meaningfully constrain $\Lambda$. As
discussed above, the comparison would have been inconclusive. This is
in accord with the fact that in the 1990s papers typically considered
several cosmological models, including $\Lambda$CDM, on a roughly
equal footing. In 1998 better data came along able to rule out the
critical-density and open models, at which point model selection would
correctly pick out the dark energy model.

b) This paragraph mentions Feynman's repackaging of all equations of
nature into $\bar{U} = 0$.  We can see no relevance in this
point. Repackaging equations does not change the number of model fit
parameters, and hence affects neither parameter estimation nor model
selection.

c) This paragraph claims that before 1992 model selection would have
argued against structure in the cosmic microwave background (CMB), on
the grounds that $C_\ell=0$ is simpler than independently specifying
each $C_\ell$.  Absolutely not! This point confuses the data and the
models. No one has ever thought that a separate specification of each
$C_\ell$ was a model, and certainly not in 1992. Indeed, clearly the
acceptable models of the time, the CDM family with or without
$\Lambda$, all predicted CMB structures that were indeed subsequently
seen. Not that this has anything much to do with model selection;
models cannot be rejected before obtaining data that actually
constrains them.

d) This paragraph notes that the galaxy two-point correlation function
was for many years thought to be a power-law, without an underpinning
physical model. However, that the power-law model is no longer
considered is irrelevant. Would any more physical model have been
wrongly ruled out by model selection, had they existed?  No is the
answer. When data improved, would model selection support them over
the power-law model? Yes. As it should.

e) This paragraph makes a point about there being the deeper physics
of the halo model behind the matter power spectrum, but this has nothing
to do with cosmological model selection.

f) This paragraph claims that the modern electro-weak theory would
have been rejected by model selection had it existed contemporaneously
with the Fermi theory during its heyday. Absolutely not!  Had the
Glashow--Weinberg--Salam model been around in say 1950, it would
\emph{not} have been ruled out by model selection because all of its
parameters were poorly constrained. Model selection would have been
unable to distinguish it from the simpler Fermi model. Later on better
data came along and ruled out the simpler model. Just as it should.

g) This paragraph makes a point that seemingly complex phenomenology
may have a simple underlying structure, e.g.~atomic spectra. There is
some relevance to this point, though a `complicated' say two-parameter
equation of state model for dark energy is unlikely to have a
substantially simpler `fundamental' description. Nevertheless, if
improved physical understanding comes along and creates a compelling
model of that type, then that is the time to try out model selection
statistics on it. Such a model can hardly be ruled out before it even
exists, nor tested until its predictions are defined.

\subsection{Linder and Miquel: Section IV}

In Section IV the authors advocate a frequentist `rejection of null
hypothesis' test where $\Lambda$CDM is the null hypothesis. This is
done by simulating data only for $\Lambda$CDM and drawing likelihood
contours, with the viability of $\Lambda$CDM then interpreted
according to the position of actual measurements with respect to those
contours. Note that this approach seeks to rule out $\Lambda$CDM in
favour of a more general dark energy model without ever computing the
probability of the data under the latter model. 

We first note that the quantity they compute and call BIC is
\emph{not} the BIC. A giveaway is that they are claiming that the
lower likelihood models are preferred. The correct computation of the
BIC requires simulation of data at each point in the parameter space,
and then a model comparison test of $\Lambda$CDM versus the
two-parameter dark energy model at each point. We carried out exactly
such an analysis, computing the full evidence rather than the BIC, in
Ref.~\cite{MPCLK}. Linder and Miquel only simulate $\Lambda$CDM, and
then simply flip the sign of the relative log-likelihood, which is not
equivalent.

On top of the above flaw, Linder and Miquel's argument then goes on to
describe a situation in which a frequentist analysis delivers a $90\%$
confidence contour around the $\Lambda$CDM model (based on synthetic
data) in the $(w_0, w_a)$ plane, and claims that a measurement lying
outside that region would exclude the $\Lambda$CDM at the $90\%$
confidence level. This is an incorrect statement in frequentist
statistics, as it slips in the wrong assumption that the probability
of the data given the hypothesis (i.e., the frequentist confidence
region) is the same as the probability of the hypothesis given the
data.\footnote{To convince oneself of the difference between the two
quantities, imagine selecting a person at random --- the person can
either be male or female (our hypothesis). If the person is female,
her probability of being pregnant (our data) is about $3\%$, i.e.\
$P(\text{pregnant}|\text{female}) = 0.03$.  However, if the person is
pregnant, her probability of being female is much larger than that,
i.e.  $P(\text{female}|\text{pregnant})\gg 0.03$. For further details,
see Ref.~\cite{lyons}.} The latter quantity is undefined for a
frequentist, for whom a hypothesis is either true or false (although
we might not know which one is true) and a probabilistic statement
about it would be meaningless. For a Bayesian of course the two are
related through Bayes theorem.

Linder and Miquel also voice their discontent about the BIC condition
being stronger, i.e.~making it harder to rule out $\Lambda$CDM. But
this is hardly surprising, being just a manifestation of Lindley's
well-known `paradox' \cite{lindley}; as summarized in Appendix~A of
Ref.~\cite{Trotta} (see Figure A1), in general frequentist
significance tests do not agree with Bayesian model selection, since
the former ignore the information gained through the data. This is
evident in Figures 1 and 2 of Ref.~\cite{MPCLK}, which is exactly the
comparison Linder and Miquel are trying to make. There is no basis to
claim, using a frequentist significance test, that the BIC `spuriously
rules out' a particular set of models, because there is no basis to
take the frequentist result as the `truth'. We could equally well say
that model-level Bayesian inference demonstrates that parameter
estimation `spuriously rules out $\Lambda$CDM' in those circumstances.

It is true that model selection gives a larger parameter area in which
$\Lambda$CDM would not be ruled out even if it is wrong, though
usually returning an inconclusive verdict in that case, to be deferred
to future data. The trade-off is that parameter estimation techniques
applied for model comparison are much more likely to rule out
$\Lambda$CDM even if it is right (by not recognizing that the data has
lower probability under a less predictive model). One cannot win on
both sides of that coin.

Independently of the above misconceptions, Linder and Miquel further
claim that if $\Lambda$CDM results in a poor likelihood in light of
new data, then it should be rejected in favour of a more general (less
predictive) model, i.e.~one in which $w_0$ and $w_a$ vary freely over
any range. It is not clear for example why {\em this} model was chosen
instead of, for example, one where $w(z)$ varies in 1000 redshift
bins, which would probably achieve an even better fit. In the Bayesian
framework we can admit all those models, assigning a prior probability
to each that reflects our relative degree of belief based on our
understanding of the physical processes at work. One then goes on to
compute the posterior probabilities for each of the models.

\subsection{Linder and Miquel: Section II}

We disagree with all statements in Section II of their paper implying
that parameter estimation and model selection are distinct
endeavours. In addition we note

1) The statement that we wouldn't want to throw away a tree containing
one fit fruit is misleading. If the fit fruit is a better fit that
those on other trees, then the goodness-of-fit will be rewarded by
model selection. If it is no fitter than those on smaller trees (and
is everywhere constrained meaningfully by data) then of course we
\emph{do} want to throw away that tree: this is what Occam's Razor is
all about and without it we have no control over arbitrarily complex
models.

2) It is implied that model selection might disadvantage fundamental
models that might have apparently complicated phenomenological
manifestations. Specific examples mentioned are braneworld models of
modified gravity and inverse power-law potentials. This criticism is
not true at all; people are welcome, indeed encouraged, to deploy
fundamental parameters in model selection rather than phenomenological
ones where possible.

\subsection{Rhetoric}

We end by pointing out that this paper uses the rhetorical trick of
attributing, without citation, and then rebutting, some vaguely
ridiculous assertions supposedly held by model selection
advocates. For instance, no-one has suggested that model selection
techniques should be `blindly applied' without regard to physical
insight, and if they had it would have been a pretty ludicrous
suggestion. No one has claimed that parameter fitting is `misguided',
it being a key part of the inference procedure, though we have indeed
argued that it is inadequate if one wishes to answer questions phrased
at the model level (e.g.~is quintessence a better description of data
than a particular modified gravity model). We are also unaware of any
cases where `overenthusiastic application of model selection led to
some claims about the probability of future experiments failing to see
characteristics such as dynamics that current data cannot access',
though we may have been enthusiastic about being able to make
probabilistic forecasts under carefully-defined prior assumptions
\cite{Trotta,PLMP,LMPW,PPOD}.

\newpage 



\begin{thebibliography}{99}
\bibitem{LM} E. V. Linder and R. Miquel, astro-ph/0702542v2.
\bibitem{revs} H. Jeffreys, {\it Theory of
    Probability}, 3rd ed, Oxford University Press (1961);
D. J. C. MacKay, {\it Information theory,
    inference and learning algorithms}, Cambridge University Press
    (2003);
P. Gregory, {\it Bayesian Logical
    Data Analysis for the Physical Sciences}, Cambridge University
    Press (2005).
\bibitem{LasHob} A. R. Liddle, P Mukherjee, and D. Parkinson, A\&G
  {\bf 47}, 4.30 (2006), astro-ph/0608184; A. Lasenby and
  M. P. Hobson, Proc. Sci. (CMB2006) 014.
\bibitem{Trotta} R. Trotta, astro-ph/0504022.
\bibitem{C95} R. D. Cousins, Am. J. Phys. {\bf 63}, 5 (1995).
\bibitem{ev} A. Jaffe, Astrophys. J. {\bf 471}, 24 (1996),
    astro-ph/9501070; P. S. Drell, T. J. Loredo, and I. Wasserman,
    Astrophys. J. {\bf 530}, 593 (2000), astro-ph/9905027; M. V. John
    and J. V. Narlikar, Phys. Rev.  D{\bf 65}, 043506 (2002),
    astro-ph/0111122; A. Slosar et al.,
    Mon. Not. Roy. Astron. Soc. {\bf 341}, L29 (2003),
    astro-ph/0212497; T. D. Saini, J. Weller, and S. L. Bridle,
    Mon. Not. Roy.  Astron. Soc. {\bf 348}, 603 (2004),
    astro-ph/0305526; P. J. Marshall, M. P. Hobson, and A. Slosar,
    Mon. Not. Roy.  Astron. Soc. {\bf 346}, 489 (2003),
    astro-ph/0307098; A. Niarchou, A. Jaffe, and L. Pogosian,
    Phys. Rev. D{\bf 69}, 063515 (2004), astro-ph/0308461;
    G. Lazarides, R. Ruiz de Austri and R. Trotta Phys. Rev. D{\bf 70}
    123527 (2004), hep-ph/0409335; M. Beltr\'an,
    J. Garcia-Bell\'{\i}do, J. Lesgourgues, A. R. Liddle, and
    A. Slosar, Phys. Rev. D{\bf 71}, 063532 (2005), astro-ph/0501477;
    P. Mukherjee, D. Parkinson, and A. R. Liddle,
    Astrophys. J. Lett. {\bf 638}, L51 (2006), astro-ph/0508461;
    M. Kunz, N. Aghanim, L. Cayon, O. Forni, A. Riazuelo, and
    J. P. Uzan, Phys. Rev. D73 (2006) 023511, astro-ph/0510164;
    M. Bridges, A. N. Lasenby, and M. P. Hobson,
    mon. Not. Roy. Astron. Soc. {\bf 369}, 1123 (2006),
    astro-ph/0511573; D. Parkinson, P. Mukherjee, and A. R. Liddle,
    Phys. Rev. D{\bf 73}, 123523 (2006), astro-ph/0605003; M. Bridges,
    A. N. Lasenby, and M. P. Hobson, astro-ph/0607404; R. Trotta,
    astro-ph/0607496; R. Trotta, Mon. Not. Roy. Astron. Soc. {\bf
    375}, L26 (2007), astro-ph/0608116; M. Kunz, B. A. Bassett, and
    R. Hlozek, astro-ph/0611004; P. Serra, A. F. Heavens, and
    A. Melchiorri, astro-ph/0701338; N. Bevis, M. Hindmarsh, M. Kunz,
    and J. Urrestilla, astro-ph/0702223; A. Niarchou and A. Jaffe,
    astro-ph/0702436; A. F. Heavens, T. D. Kitchling, and L. Verde,
    astro-ph/0703191
\bibitem{BCK} B. A. Bassett, P. S. Corasaniti and M. Kunz,
    Astrophys. J. Lett. {\bf 617}, L1 (2004), astro-ph/0407364.
\bibitem{MPCLK} P. Mukherjee, D. Parkinson, P. S. Corasaniti,
    A. R. Liddle, and M. Kunz, Mon. Not. Roy. Astron. Soc. {\bf
    369}, 1725 (2006), astro-ph/0512484.
\bibitem{KTP} M. Kunz, R. Trotta and D. Parkinson,
        Phys. Rev. D {\bf 74} 023503 (2006), astro-ph/0602378.
\bibitem{PLMP} C. Pahud, A. R Liddle, P. Mukherjee, and D. Parkinson,
  Phys. Rev D{\bf 73}, 083523 (2006), astro-ph/0605004.
\bibitem{LMPW} A. R. Liddle, P. Mukherjee, D. Parkinson, and Y. Wang,
  Phys. Rev. D{\bf 74}, 123506 (2006), astro-ph/0610126.
\bibitem{PPOD} R. Trotta, astro-ph/0703063.
\bibitem{S78} G. Schwarz, Ann. Statist. {\bf 5}, 461 (1978).
\bibitem{L04} A. R. Liddle, Mon. Not. Roy. Astron. Soc. {\bf 351},
    L49, astro-ph/0401198; A. R. Liddle, astro-ph/0701113.
\bibitem{lindley} D. Lindley, Biometrika {\bf 44}, 187 (1957).
\bibitem{lyons} L. Lyons, `A particle physicist's perspective
on astrostatistics', to appear in {\em Proceedings of the
conference `Statistical Challenges in Modern Astronomy IV'}, State
College, Pennsylvania, USA, June 2006.
\end{thebibliography}
\end{document}